\documentclass[11pt]{article}
\usepackage{coling2020}
\usepackage{times}
\usepackage{url}
\usepackage{latexsym}
\usepackage{graphicx}
\usepackage{algorithm}
\usepackage{algpseudocode}

\colingfinalcopy 

\title{KOSMOS: Knowledge-graph Oriented Social media and Mainstream media Overview System}

\author{\underline{Chua} Hao Yang, \underline{Yong} Shan Jie \\ \\
  Nanyang Technological University \\ Singapore \\
  {\tt \{chua0808,yo0001ie\}} \\
  {\tt @e.ntu.edu.sg} \\\And
  \underline{Boon} Kok Chin,  Lander \underline{Chin},\\
  \textbf{Lynnette Hui Xian \underline{Ng}} \\
  Defence Science Technology Agency \\
  Singapore \\
  {\tt\{boonkc,lanchin,nhuixia1\}} \\
  {\tt @dsta.gov.sg}  \\
}

\date{}

\begin{document}
\maketitle
\begin{abstract}
We introduce KOSMOS, a knowledge retrieval system based on the constructed knowledge graph of Social media and Mainstream media documents. The system first identifies key events from the documents at each time frame through clustering, extracting a document to represent each cluster, then describing the document in terms of 5W1H (Who, What, When, Where, Why, How). The event-centric knowledge graph is enhanced by relation triplets and entity disambiguation from the representative document. Knowledge retrieval is supported by a web interface that presents a graph visualisation of related nodes and relevant articles based on a user query. The interface facilitates understanding relationships between events reported in mainstream and social media journalism through the KOSMOS information extraction pipeline, which is valuable to public office to understand media slant and public opinions. Finally, we explore a use case in extracting events and relations from documents to understand the media and community's view to the 2020 COVID19 pandemic. 

\end{abstract}

\section{Introduction}
\label{intro}
Understanding relationships between events reported in mainstream news articles and social media chatter is a key goal of information extraction, aiding policy makers to sense make the vast information space and understand citizen chatter on specific issues. 

We present KOSMOS, a web interface to understand online information through an event-centric knowledge graph. We draw our data from mainstream news articles and social media articles, henceforth collectively referring to the data as documents. This paper presents the information processing pipeline through identifying salient events (Section \ref{sec:identifyingthemes}), relation extraction and knowledge graph construction (Section \ref{sec:knowledgegraphconstruction}), and knowledge retrieval (Section \ref{sec:knowledgeretrieval}). Finally, we explore a use case in extracting events and relations from documents to understand the media and community's view to the 2020 COVID19 pandemic (Section \ref{ref:usecase}).



\section{Related Works}
KOSMOS is made up of several key components in its pipeline; Data Scraping, Events Identification, Knowledge Graph, and Knowledge Retrieval. Some related work would include NIFTY \cite{DBLP:conf/www/SuenHESL13}, which performs clustering of topics from various sources across the internet by partitioning phrases from document, then presenting each cluster as a Directed Acyclic Graph. VisIRR, \cite{7042511} seeks to ease information retrieval and recommendation through interactive scatter plots. 

Besides identifying key themes in the documents, the next goal is to turn unstructured information from documents into a structured form of relation tuples to express the relationship between entities and by extension, documents. Several approaches have been developed to construct knowledge bases using news data \cite{ROSPOCHER2016132} \cite{10.1145/3308560.3316761} and social media data \cite{gottschalk2019eventkg}, representing events, entities and relations between the data. Many query methods include BERT for natural language understanding \cite{DBLP:conf/iclr/DhingraZBNSC20}, followed by knowledge graph embedding \cite{10.1145/3289600.3290956} to find the relevant graph nodes by breaking Natural Language Queries (NLQ) into its entities and predicates. 

Most existing solutions build knowledge graphs based on static Wikipedia databases or structured QA data sets such as DBpedia and Wikidata, focusing on a single type of document structure, and presenting results in one single presentation view. The aim of KOSMOS is to put together the classification of multi-source information, as well as to provide an intuitive information retrieval interface, displaying what is relevant in a graphical and temporal representation, to enhance sense making through finding entities related, and how sequence of events unfold.

\textbf{Front-End and Server Frameworks.} To implement KOSMOS, we relied on several frameworks and libraries to implement and integrate the full pipeline. Start from the front-end, we used JavaScript React for how it renders changes very rapidly with its Virtual DOM changing only what has been updated. We coupled the front end with a web server. For the web server, we chose Python Flask over other framework such as Django because of its lightweight interface and modularity, allowing us to build the back end server linking up all other components through REST API. 

\textbf{Libraries for Data Processing.} On the backend, there are a couple more libraries used to extract crucial information from unstructured text. Firstly, to perform document clustering, spaCy's word vector \cite{spacy2} was used to vectorise the document. spaCy's library was used for this task because of its faster computational time compared to Stanford NLP's GloVe library. However, For relation triples $<$subject-relation-object$>$ extraction and recognition of named entity, we used Stanford NLP's OpenIE extractor \cite{angeli-etal-2015-leveraging} and Named Entity Recognizer \cite{ner} because spaCy offered only entity extraction without any identifying relations between extracted entities.

Each document that enters the KOSMOS pipeline needs to have a concise breakdown of the document at the end of its processing. A candidate for this task would be Bert Extractive Summarizer \cite{miller2019leveraging}. However, text summarizers, despite it returning a concise summary of the document, is still largely unstructured. We thus adopted the Giveme5W1H library \cite{Hamborg2019b}, which takes the document and answers the 6 general questions - Who, What, When, Why, How - about the text.

\textbf{Database and Data Retrieval Frameworks.} Two form of database is required for KOSMOS - Graph Database and Document Database. Neo4j was the choice for Graph Database over others because Neo4j allowed for properties to be stored within nodes, unlike other Graph Database, such as Ontotext's GraphDB, which primarily adopts the RDF store database model, which represents each property of a node as additional node and edge. This perks of Neo4j would mean only entities are made into nodes, and properties are stored within the created entities, allowing the output graph to be significantly more condensed. On top of that, Neo4j also has a Desktop application for us to verify our graph with great ease.

Lastly, for a Document Database, we need a framework that not only ensures data integrity, but also able to retrieve documents with ease and rapidly. For KOSMOS, we used Elasticsearch as a NoSQL database. Elasticsearch is a highly scalable and fast textual-based search engine on top of being a database. It has many useful features to enhance our document retrieval, such as multi-field matching, and fuzzy search.

\section{System Overview}
\begin{figure*}[h]
\centering
\includegraphics[width=0.8\textwidth]{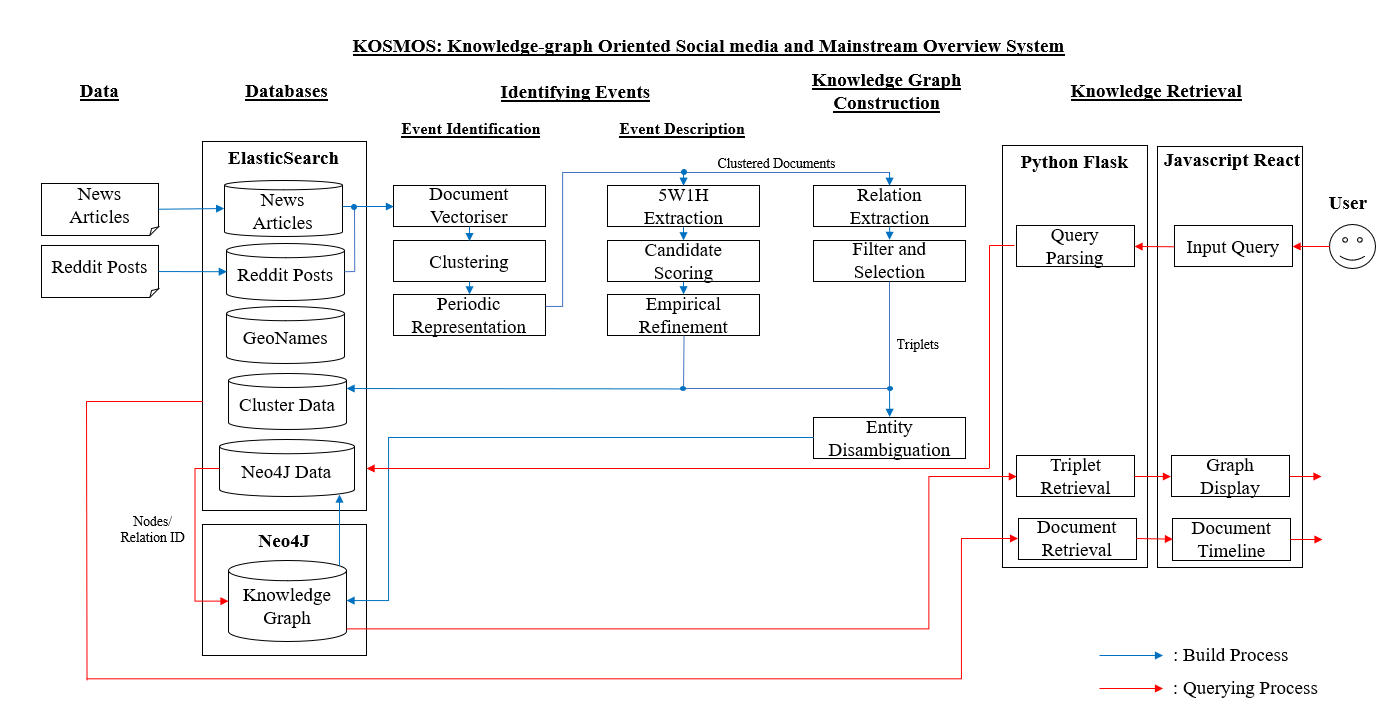}
\caption{KOSMOS System Overview}
\label{fig:pipeline-fig}
\end{figure*}

Figure \ref{fig:pipeline-fig} presents an overview of the pipeline of the KOSMOS system. We collected articles from mainstream news sources using RSS feeds and Reddit posts using Pushshift.io \cite{baumgartner2020pushshift}. The collected data are stored in an ElasticSearch database, which facilitates document search with an expandable schema. The knowledge graph stores entity relational data in a Neo4J database. The pipeline is implemented in Python, while the user interface for the Knowledge Retrieval module is implemented with Javascript React and supported by a Python Flask server. 

The pipeline first identifies salient events by performing document clustering to group documents across time periods, then identify a representative document that represents the theme of the cluster. To further describe the event, event descriptors of 5W1H (Who, What, When, Where, Why, How) are extracted on the representative document. The knowledge graph is constructed in the next step from all the representative documents. This is done by extracting relations through identifying relation triplets and performing entity disambiguation before forming nodes and links in the graph database. A user interface facilitates information retrieval of the relationships between documents and entities related to the user query.

\subsection{Identifying Events in Documents}
\label{sec:identifyingthemes}
At each time period, a large amount of documents enter the pipeline. Many mainstream articles and Reddit posts may elaborate on the same event, but a handful of Reddit posts are incoherent sentences. To remove these noise and identify salient news events, we first perform document clustering, which results in clusters of documents and further highlights one representative document per cluster. Each event in a cluster is processed using Hamborg's 5W1H library to extract a more concise event descriptors. 

\textbf{Document Clustering.}
The documents are first pre-processed by stop-word and non-ASCII character removal, deduplication, tokenisation and lemmatisation. Non-English documents are disregarded to focus on an English-based system.



Each document is vectorised into 300 dimensions by sum average using spaCy's word vector \cite{spacy2}. To pick out salient features in the documents, we reduce the documents to 100 dimensions by Principal Component Analysis, giving a variance ratio of 0.92 and 0.84 on news and Reddit documents respectively, intuitively representing the amount of initial semantics in each document type.

Density-based spatial clustering is then applied on the vector collection with a cosine distance metric. We perform a parameter search to find the optimal distance between two documents as the distance resulting in the maximum number of clusters. To keep the documents in each cluster concise and consistent, each cluster is represented by one representative document. This representative document is determined by finding the one closest to the cluster center.




\textbf{Extracting 5W1H.} For the representative document in each cluster, the system extracts the main event descriptors in the form of 5W1H from the article using the library Giveme5W1H \cite{Hamborg2019b}. We find that refining the search results to retain the top two descriptors based on the probability of confidence gives a more reasonable and desired result than its default, based on empirical testing on our articles. The extraction of event descriptors for each document cluster provides a quick summary of the events at each time period.



\subsection{Knowledge Graph Construction.}
\label{sec:knowledgegraphconstruction}
We enhance the knowledge graph to capture entity relationships from the representative document extracted in the previous section. This knowledge graph is constructed using Neo4J, which serves as a base for extracting information with the aid of a user search module. 


\textbf{Relation extraction.} Stanford NLP's libraries are used to perform relation extraction on the representative document. Relation triples are extracted from the article's text using OpenIE extractor \cite{angeli-etal-2015-leveraging}. The triplets are then filtered using the Named Entity Recognizer \cite{ner} to keep essential entities to a knowledge graph and trimmed to eliminate duplicated triples, before performing entity disambiguation against the knowledge graph.



\begin{figure*}[h]
\centering
\includegraphics[width=0.8\textwidth]{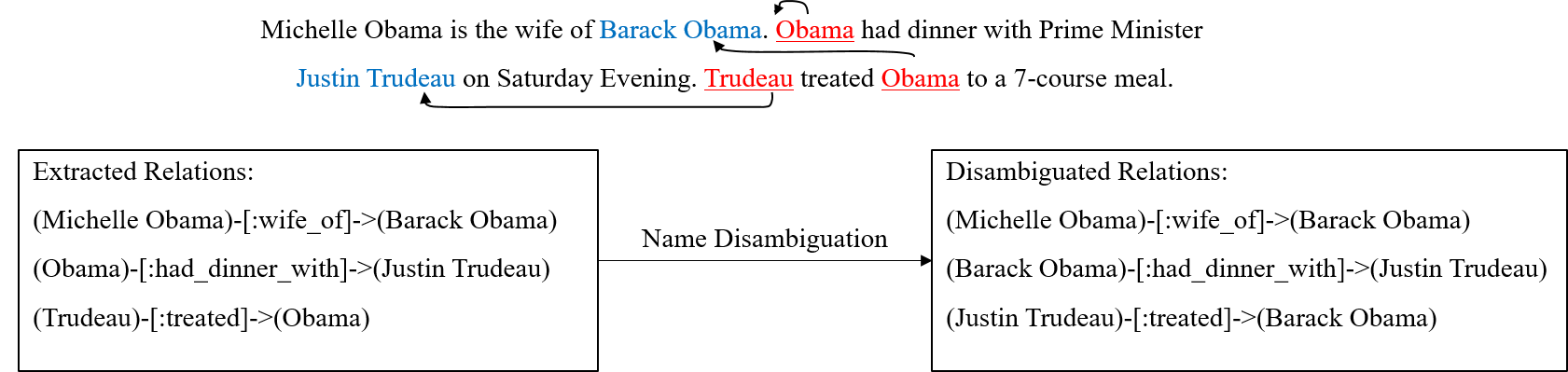}
\caption{Named Entity Disambiguation Workflow}
\label{fig:entitydisambiguation}
\end{figure*}

\textbf{Entity Disambiguation.}
\label{sec:entitydisambiguation} 
In insertion of extracted relations into the Neo4J database, entity disambiguation is required as many news articles have similar relations and/or entities. In this step, we perform deduplication of triplets with the existing knowledge base and addition of nodes and corresponding links to existing nodes. We harness the GeoNames database\cite{geonames} to map the geographical information from extracted location nodes to make geographical sense of these nodes by tying respective cities to their countries. In addition, we eliminate the creation of separate nodes from the mention of last names with full names, presented in Figure \ref{fig:entitydisambiguation}. The resulting set of triplets are then inserted into the knowledge graph. The edge between nodes stores the information of their respective document's data. The extracted entities are then tokenised and the tokens stored in the ElasticSearch database to facilitate retrieval of nodes during query of knowledge retrieval module.


\subsection{Knowledge Retrieval}
\label{sec:knowledgeretrieval}
KOSMOS knowledge retrieval module presents a simple to access web browser based interface built using Javascript React library and Python Flask. Multiple users can access and perform search queries at the same time. The application facilitates a user search query, then displays a graph of related entities to the query retrieved from the Neo4J database, uses a timeline to display the documents.


\textbf{Deconstructing the User Query}
To facilitate the search through the knowledge base, we built a Natural Language Query (NLQ) search to translate the user's query into the corresponding ElasticSearch Lucene Language, followed by Neo4J Cypher language to retrieve the relevant information. While there are query methods using graph vertices comparison \cite{Truong2008GVCAG}, we found that leveraging on the Cypher and Lucene search language produces acceptable results. The user inputs a query from the search box, which is then passed to the ElasticSearch database to retrieve relevant entity/relation's ids by leveraging on the ability of the Lucene query syntax to perform multi-match queries across tokens.

\textbf{Retrieving Knowledge Graph from Neo4J.}
After the ElasticSearch query returns the relevant node and relation id, a Neo4J Cypher query is constructed to retrieve the nodes and the first degree relations from the knowledge graph. This knowledge association is displayed to the user using the ReactForceGraph Javascript library.

\textbf{Retrieving Articles from ElasticSearch.} Relationships between Nodes are formed from article clusters, with each cluster represented by a representative document (Section \ref{sec:knowledgegraphconstruction}). The corresponding cluster's representative document is displayed upon clicking a relation edge, and one may drill down to individual documents. We note that there are many documents that do not fit into any cluster, and have elected to present the documents alongside the clustered articles, leveraging on the Lucene query language to search through the document database to provide a comprehensive overview of the event.

\section{Use Case Demonstration}
\label{ref:usecase}
We collected documents during the period of January to March 2020, mapping documents related to the COVID19 world pandemic. We currently have 13,709 news articles collected from mainstream sources, and 36860 Reddit posts from r/coronavirus forming 158 clusters of documents. This forms 5525 nodes and 5441 relationships. We characterised clusters by days to keep up with the COVID19 pandemic news.

\begin{figure*}[h]
\centering
\includegraphics[width=0.8\textwidth]{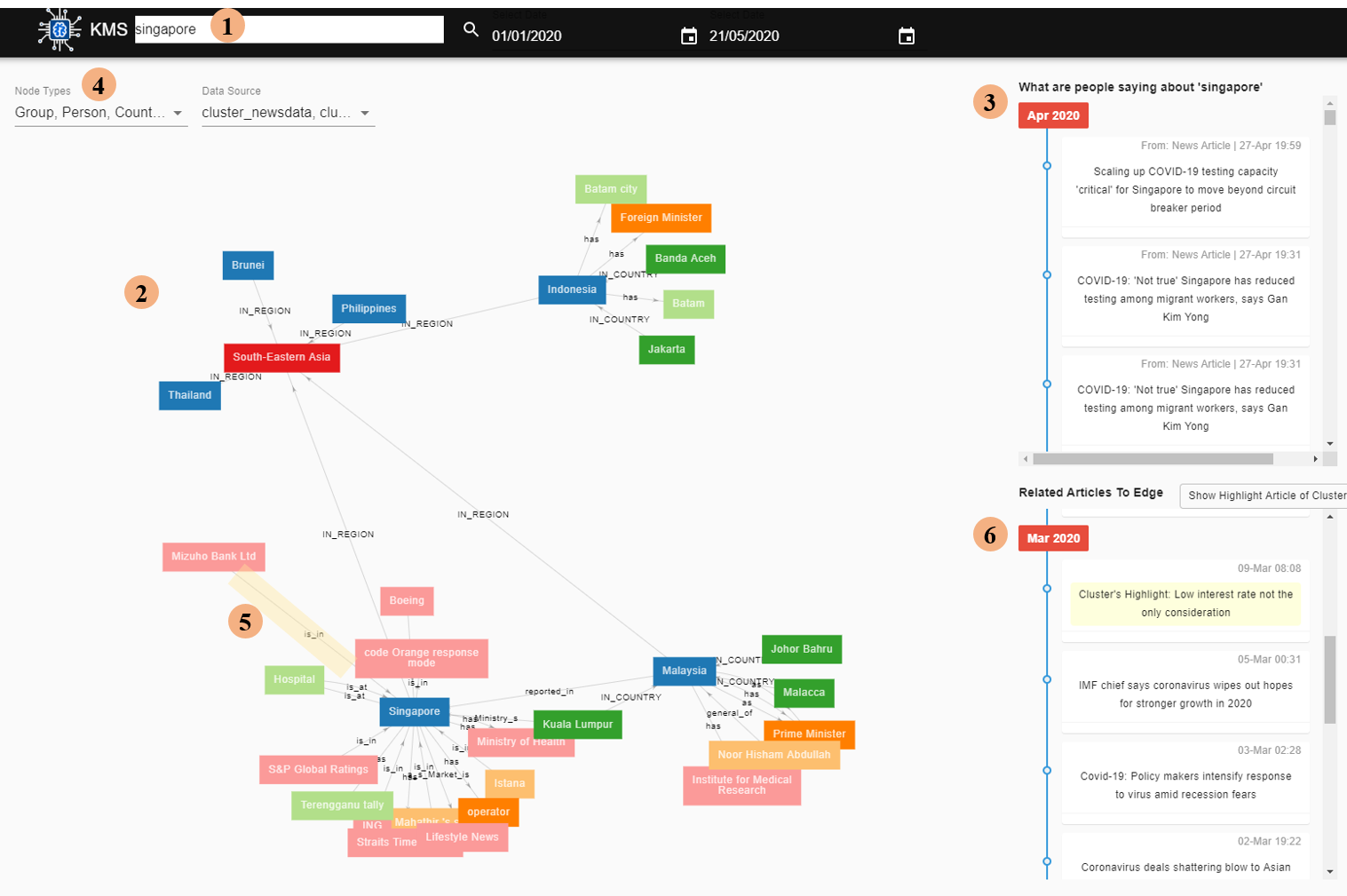}
\caption{KOSMOS system presenting relations and documents to 'Singapore'}
\label{fig:screenshot1}
\end{figure*}

Figure \ref{fig:screenshot1} showcases the KOSMOS user interface. A knowledge search begins with \textbf{(1)}, where the user inputs a search query, which brings up a knowledge graph  of entities that match to the search query and the first degree relationships \textbf{(2)}. A timeline displays  retrieved articles matched through ElasticSearch query feature \textbf{(3)}. The presented graph have been filtered (through \textbf{(4)} node type and data source selection). The edge \textbf{(5)} is expanded, where a timeline of related articles is displayed on \textbf{(6)}.

\section{Conclusion and Future Work}
We have presented KOSMOS as the first event-centric knowledge graph information retrieval system, dynamically constructed from news and social media documents. Our system fuses document clustering and knowledge graph construction to understand relationships between documents. Through identifying key document themes via document clustering, the noise in the data is reduced, and with it, and the system's existing knowledge graph is enhanced with salient entities and relationships. Based on the knowledge graph, KOSMOS delivers an overview of events through knowledge retrieval of nodes and documents, allowing easier sensemaking through the vast information space. This work naturally entails a couple of interesting direction for future research: (1) enhancing the breadth of information through incorporation of more mainstream and social media sites; (2) incorporating a question-and-answer query system to parse natural language queries. 





\bibliographystyle{coling}
\bibliography{coling2020}

\end{document}